\documentclass[10pt, conference, letterpaper]{IEEEtran}

\usepackage{cite}
\usepackage{subfigure}
\usepackage{url}
\usepackage{verbatim,epsf}
\usepackage[caption=false,font=footnotesize]{subfig}
\usepackage{enumitem}

\usepackage{tabularx}
\usepackage{balance} %MAU this is needed to balance the column of the last page

\usepackage{amssymb}
\usepackage{amsmath}
\usepackage{graphicx}
\usepackage{epstopdf}

\usepackage{epsfig, color}

\usepackage{comment} % Package per i commenti
\usepackage{amsfonts} % Package per i caratteri  matematici spciali
\usepackage{amsmath} % Package per i caratteri  matematici spciali
\usepackage{rotating}

\usepackage{amssymb}
\usepackage{amsmath}
\usepackage{amsfonts}
\usepackage{wasysym}

\usepackage{algorithm}
\usepackage{algorithmic}

\newsavebox{\ieeealgbox}

\newlength\myindent
\IEEEaftertitletext{\vspace{-0.28in}}

\begin{document}

\title{%\huge
	{A Game-theoretic Framework for Revenue Sharing in Edge-Cloud Computing System}}
 	%Incentive Mechanism for Edge Computing via Shapley-value} 
\author{ 
	\IEEEauthorblockN{Zhi Cao\IEEEauthorrefmark{1}, Honggang Zhang\IEEEauthorrefmark{1},
		Benyuan Liu\IEEEauthorrefmark{2}, Bo Sheng\IEEEauthorrefmark{1}\\
		\IEEEauthorrefmark{1} UMass Boston, Boston, MA. Email: \{zhi.cao001,honggang.zhang,bo.sheng\}@umb.edu \\
		\IEEEauthorrefmark{2} UMass Lowell, Lowell, MA. Email: bliu@cs.uml.edu
	}
}

\maketitle
\pagestyle{plain}

\begin{abstract}
	
We introduce a game-theoretic framework to explore revenue sharing in 
an Edge-Cloud computing system, in which computing service providers
at the edge of the Internet (\textit{edge providers}) and 
computing service providers at the cloud (\textit{cloud providers})
co-exist and collectively provide computing resources to clients (e.g., end users or applications)
at the edge. Different from traditional cloud computing, 
the providers in an Edge-Cloud system are independent and self-interested. 
To achieve high system-level efficiency, the manager of the system adopts a task distribution
mechanism to maximize the total revenue received from clients and also adopts a 
revenue sharing mechanism to split the received revenue among computing servers (and hence service providers). 
Under those system-level mechanisms, service providers attempt to game with the system 
in order to maximize their own utilities, by strategically allocating their resources (e.g., computing servers).

Our framework models the competition among the providers in an Edge-Cloud system 
as a non-cooperative game. 
Our simulations and experiments on an emulation system have shown 
the existence of Nash equilibrium in such a game.
We find that revenue sharing mechanisms have a significant impact on the 
system-level efficiency at Nash equilibria, 
and surprisingly the revenue sharing mechanism based directly on actual contributions can result 
in significantly worse system efficiency than Shapley value sharing mechanism 
and Ortmann proportional sharing mechanism. 
Our framework provides an effective economics approach to 
understanding and designing efficient Edge-Cloud computing systems.

\end{abstract}

\section{Introduction}

Edge computing \cite{bonomi2012fog,chiang2016fog,hu2015mobile,shi2016edge,yi2015survey,liu2016paradrop,yoon} 
is an emerging computing paradigm that is transforming 
the landscape of provision and consumption of computing services for a wide range of applications
and end users at the edge of the Internet. This paradigm will be particularly helpful to
those latency-sensitive and bandwidth-hungry (due to a large amount of data) 
applications brought by Internet of Things (IoT) systems. 

In this paper, we are interested in a hybrid system where 
computing service providers at the edge of the Internet (referred to as \textit{edge providers}, 
which are close to IoT sensors, mobile devices, and end users)  
and the providers at the cloud (referred to as \textit{cloud providers}) co-exist and collectively 
provide computing services to the clients at the edge. Such a system is referred to as an \textit{Edge-Cloud system}. 
Different from a traditional cloud computing environment in which all servers are organized in data centers
and tightly controlled and managed by a provider, 
the various providers that offer computing servers in an Edge-Cloud system are independent 
and located at various distances away from clients,
and they can make independent decisions on the computation resources that they provide to the system.  

In order to achieve a high system-level efficiency, an Edge-Cloud system 
adopts a task distribution mechanism to maximize the total revenue received from clients, 
and it adopts a revenue sharing mechanism to fairly split its received revenue 
among computing servers (and hence service providers).
Under any given system-level mechanism, the providers are likely to compete with each other and game with the system
in order to maximize their own utilities by strategically adjusting the resources
that they offer to the system.

The self-interested behaviors of those providers might result in an inefficient system with low overall performance,
as their individual self-interested objectives (a provider tries to maximize its 
own utility) do not collectively align with the system-wide objective (i.e., to maximize the overall system utility).
Therefore, it is important to choose a system-level mechanism that will minimize 
the loss of system-wide efficiency in the face of the self-interested behaviors of the providers.  
To that end, we introduce a game-theoretic framework to investigate 
the impact of revenue sharing mechanisms on an Edge-Cloud system's overall efficiency. 
To demonstrate the effectiveness of the framework, 
we have conducted extensive simulations and experiments on an edge-computing emulation system.

Our major contributions are summarized below. 
\begin{enumerate}
	\item We introduce a game-theoretic framework to investigate 
	an Edge-Cloud hybrid computing system of edge providers  and 
	cloud providers offering their computing resources to clients
	at the edge. 
	To the best of our knowledge, our work is the first to investigate such a problem
	from an economics and game-theoretic perspective.
	Our findings demonstrate that it is crucially important to design an appropriate revenue sharing mechanism 
	in order to maintain high system-level efficiency in the face of service providers' self-interested behaviors.

	\item 
	We demonstrate the existence of Nash equilibrium in the game 
	between edge and cloud providers, under three revenue sharing mechanisms (Shapley value sharing \cite{shapley},  
	Ortmann proportional sharing \cite{ortmann2000proportional}, and Direct-contribution-based sharing), 
	and across a wide range of system/networking settings.
	We find that when the servers from different providers have quite different capacities (e.g., the transmission
	bandwidth between clients and servers), 
	different revenue sharing mechanisms can result in drastically different 
	system-level utility loss at equilibria when compared with maximum system utility (which is
	achieved when providers do not game with the system).
	
	\item Our results show that at the Nash equilibria of the game,
	Direct-contribution-based sharing (i.e., revenue split based directly on actual contributions of servers) 
	results in the worst system-level utility. 
	This seemingly counter-intuitive result is not surprising, 
	as under Direct-contribution-based sharing, a provider with very low transmission bandwidth 
	keeps placing many servers in the system (as it is rewarded directly based on its actual contribution) even when 
	doing so actually hurts the overall system performance.  
	On the other hand, Shapley mechanism gives the least utility loss in most cases, and Ortmann mechanism's utility loss 
	is close to Shapley's.
	This is because Shapley mechanism and Ortmann mechanism discourage a low bandwidth provider
	from offering many servers by setting its revenue share as a decreasing function of its number of servers placed in the system.
	
	\item We demonstrate that our framework is a valid and effective 
	economics approach to understanding and designing efficient Edge-Cloud computing systems, 
	based on our extensive simulations driven by the empirical data 
	derived from our experiments on an emulation system we have developed
	and from Google cloud trace \cite{googleclouddata}. 
		
\end{enumerate}

In the rest of the paper, we first present the architecture of an Edge-Cloud system and give an overview
of our game-theoretic framework in Section \ref{sec_ec}. Then in Sections \ref{sec_job_distribution} and 
\ref{sec_sharing}, we describe task distribution mechanisms and revenue sharing mechanisms. 
Section \ref{sec_exp} presents our findings via experiments and simulations, 
and we conclude the paper in Section \ref{sec_conclusion}.

\section{Edge-Cloud system}\label{sec_ec}

In this section, we first discuss background and related work, and then we 
give an overview of an Edge-Cloud system. 

\subsection{Background and Related Work}
This paper studies a computing system in the emerging edge computing paradigm, which 
broadly includes cloudlets, mobile edge computing, 
fog computing, fog networks, and mobile cloud computing  \cite{bonomi2012fog,chiang2016fog,hu2015mobile,shi2016edge,yi2015survey,liu2016paradrop,yoon}.  
Besides the low latency benefit of edge computing,
recent research on Internet of Things (IoT) has 
shown that, by processing at the edge 
the large amount of raw data collected by IoT sensors (e.g., in a smart home or smart city system)
or human users (e.g., videos, pictures taken by smartphones),
edge computing can significantly reduce the consumption of network bandwidth in wide area and core networks
when compared with transferring raw data to a cloud \cite{bahl2015design,ha2014towards}.  
AT$\&$T, a large Internet service provider, has recently announced plans to deploy edge computing servers 
in their mobile access towers on a large scale \cite{att2017edge}.
The economics and game-theoretic approach adopted in this paper 
is related to the existing rich literature of applying economics and game theory in  
networking research \cite{ma2010internet,misra2010incentivizing,zhang2005tcpconnection_icnp,zhang2016incentive}.

\subsection{System Architecture Overview}

\begin{figure}[htb!]
	\centerline{
		\begin{minipage}{2.5in}
			\begin{center}
				\setlength{\epsfxsize}{2.5in}
				\epsffile{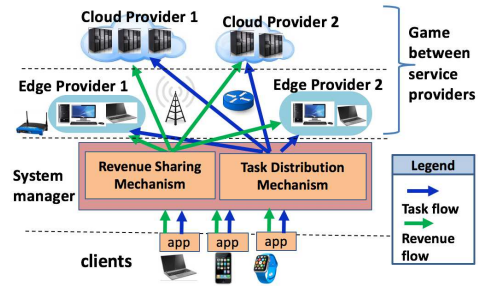}\\
				{}
			\end{center}
		\end{minipage}
	}
	\caption{System architecture. Computing service providers compete with each other under
		system-level mechanisms: task distribution and revenue sharing.}
	\label{fig_arch}
\end{figure}

There are three types of entities in an Edge-Cloud system, as shown in Figure \ref{fig_arch}.
(1) Clients, including applications and end users, 
	that are at the edge of the Internet and submit computing tasks to the system.
(2) Edge providers (computing service providers at the edge of the Internet and close to clients,
	and hence have high communication bandwidth and low propagation delay to clients), 
	and cloud providers (providers in the cloud that offer servers to edge clients by joining an Edge-Cloud system). 
(3) A system manager, which is a software component that implements mechanisms/algorithms
	for various management issues such as facilitating task submissions, revenue collection 
	from clients, revenue split among servers, accounting/auditing, etc. The main part of the 
	manager resides on the edge and some of its components are distributed among providers throughout the Internet.
In the system, clients communicate with and submit their tasks to the system manager through apps on their devices.

In an Edge-Cloud system, a monetary value is associated with each task. 
A system manager's objective 
is to maximize its total values or \textit{revenue} received from clients
via a task distribution mechanism that
optimally assigns tasks to servers subject to the latency requirements of those tasks.
Based on the revenue collected and the tasks completed by the servers,
the manager utilizes a revenue sharing mechanism
to split the received revenue among the servers (and hence between 
the service providers who own those servers). 
Therefore, an Edge-Cloud system has two basic types of mechanisms: 
a task distribution mechanism and a revenue sharing mechanism.
In this paper, we investigate three types of revenue sharing mechanisms: 
Shapley value sharing \cite{shapley}, Ortmann proportional sharing \cite{ortmann2000proportional}, 
and Direct-contribution-based sharing.
They will be discussed in detail in Sections \ref{sec_job_distribution} and \ref{sec_sharing}.

\subsection{A Game-theoretic Framework for an Edge-Cloud System}

\subsubsection{Assumptions}

We assume that each service provider in an Edge-Cloud system 
is independent and self-interested. This assumption describes an important characteristic of
an Edge-Cloud system: a service provider can choose to join 
an Edge-Cloud system and decides by itself the amount of computing resources it  
offers to the system. This characteristic differentiates an Edge-Cloud system from a traditional 
cloud computing system in which all computing resources 
are centrally managed and tightly controlled
by an entity and they are typically placed in data centers.  
A traditional cloud computing provider can also join an Edge-Cloud system 
and offer service to the clients in the system, and such a cloud provider 
is treated equally as other edge providers in the same system.

We assume that the two types of mechanisms (i.e., task distribution and revenue sharing)
work on individual servers, without any consideration of the identities of the owners (i.e., providers) 
of those servers, as the objectives of those system-level mechanisms are to optimally utilize
available \textit{servers} to maximize total received revenue and fairly distribute the revenue\footnote{A system manager 
	may keep a share of the total received revenue and split the rest among servers. We assume that  
	a system manager's own revenue share is negligible compared with the rest of the revenue given to servers.} 
among participating \textit{servers}. 
Those mechanisms are publicly known to all clients and service providers.
Under those mechanisms, a service provider attempts to maximize its received 
benefit or utility (defined below) by strategically adjusting the computing resources it provides to the system.

\subsubsection{The game} We model the competition among service providers in an Edge-Cloud system as 
a non-cooperative game \cite{basar98game}, and the providers are \textit{players} in the game. 
We focus on the case where a provider's available strategy is to adjust
the number of servers it offers to the system in order to maximize its utility.  
A utility function captures the tradeoff between the revenue and the cost of a provider. 
Providing more servers will incur more cost to a provider, 
even though more servers imply more revenue that the provider can potentially receive.
Then the utility function of a provider $p$ can be described as 
\begin{equation}
U_p(n_p) = v_{p}(n_{p}) - f_{cost}(n_{p}) \label{eqn_u_general}
\end{equation}
where 
$v_p(n_p)$ is the revenue received by provider $p$ when placing $n_p$ servers in the system, and  
the cost $f_{cost}$ is an increasing function of the number of servers. 
We focus on a linear cost function $f_{cost}(n_{p})=\alpha_p n_p$ with $\alpha_p>0$.

Now we define a Nash equilibrium \cite{basar98game} for the game we study through a game of two players:
an edge provider $E$ and a cloud provider $C$.
Let $v_{E}$ and $v_{C}$ denote the revenue received by the edge 
player and the cloud player respectively. 
Let $M_{opt}$ and $M_{share}$ denote a task distribution mechanism and a revenue sharing mechanism
respectively.
Then we know $v_{E}$ is a function of $M_{share}$ and $M_{opt}$.
Note that $M_{opt}$ is a function of $(n_{E}, n_{C})$, i.e., 
the number of servers provided by the edge player and the cloud player respectively, 
given that they work on a certain set of tasks. 

The edge player and the cloud player attempt to solve the following optimization problems respectively
\begin{eqnarray}
\max_{n_{E}} U_{E}(n_{E}, n_{C})&=& v_{E}(M_{share}(M_{opt}(n_{E}, n_{C})) )  - \alpha_{E} n_{E} \nonumber\\
\max_{n_{C}} U_{C}(n_{C}, n_{E})&=& v_{C}(M_{share}(M_{opt}(n_{E}, n_{C})) )  - \alpha_{C} n_{C} \nonumber
\end{eqnarray}

A Nash equilibrium \cite{basar98game} of the game is 
a particular combination of players' strategies from which a player has no incentive to unilaterally deviate
% from it's current strategy 
(i.e., does not change its number of servers, given that the other player's number of servers remains unchanged), 
as any unilateral deviation will not increase its utility. The Nash equilibrium is denoted by 
\begin{equation}
\{n^*_{E},  n^*_{C}\}\label{eqn_ne}
\end{equation}
where $n^*_{E}=\mbox{argmax}_{n_E} U_{E}(n_E, n^*_{C})$ and 
$n^*_{C}=\mbox{argmax}_{n_C} U_{C}(n_C,n^*_{E})$.
Note that the definition (\ref{eqn_ne}) can be easily generalized
to the definition of a Nash equilibrium of a $m$-player game:
$\{n^*_i\}, \forall i \in \{1,2,...,m\}$ with $n^*_i=\mbox{argmax}_{n_i} U_i (n_i, n^*_{-i})$ (where $-i$ denotes the set of all 
players except $i$).

\section{Mechanism for Distributing Computing Tasks}\label{sec_job_distribution}

In this section, we discuss mechanisms that distribute tasks of edge clients
to the servers in an Edge-Cloud system.

\subsection{Objective of task distribution}

Since an important application of edge computing is to serve tasks with low latency requirement, 
we focus on tasks with completion deadlines. Recall that a task has a value,
which can be regarded as the payment that the task's owner (i.e., a client) is willing to pay for completing the task. 
\textit{The objective of a task distribution mechanism 
is to maximize the total received value (as a revenue) for the tasks that are completed before their deadlines}. 
In this section, we present an optimization formulation for the case where tasks arrive in a batch (i.e., at the same time) 
to illustrate the characteristic of task distribution, 
and then we will present a greedy algorithm to address a practical dynamic task arrival setting.

\subsection{Optimal Task Distribution Formulation}\label{sec_batch}

\subsubsection{Batch task arrival} 
A system manager distributes all tasks arriving in a batch 
to servers by solving an optimization problem to maximize the total received revenue
from those completed tasks.  
We assume that the execution order of those tasks on a server should be the same
as the order of their arrivals. 
We also assume that tasks are not splittable. 

Let $\mathbb{N}_J$ denote the set of tasks with $N_J=|\mathbb{N}_J|$, 
and let $\mathbb{N}_S$ denote the set of servers with $N_S=|\mathbb{N}_S|$.  
Tasks are ordered increasingly according to their arrival times and indexed by $i=1, ..., N_J$, 
and servers are indexed by $j=1,..., N_S$.
Let $x_{ij}$ denote the assignment of task $i$ to server $j$.
Then $x_{ij}=1$ represents that task $i$ is assigned to server $j$; otherwise $x_{ij} = 0$.
Let $d_{ij}$ denote the completion time of task $i$ when it is assigned to server $j$. 
Note that $d_{ij}$ includes the computation time of task $i$ on server $j$ and the time to transfer task $i$ 
to server $j$. In addition, a task $i$ might experience a queuing delay 
if some other tasks are scheduled on the same sever (as task $i$) but should be executed before task $i$
as they arrive earlier than task $i$.
Queuing delay is discussed next.

Let $v_i$ denote the value of task $i$ or the payment 
that the owner (i.e., client) of task $i$ will pay for completing task $i$.
If task $i$ is completed before its deadline, the system manager 
will receive $v_i$; otherwise, the manager receives nothing.  
The objective of the manager is to 
maximize its total received payment or value (as a revenue) by solving the following optimization problem. 
\begin{eqnarray}
& \max_{x_{ij}} \quad \sum_{j=1}^{N_S} \sum_{i=1}^{N_J} v_{i} x_{ij}& \label{val_max}\\
s. t. & 0 \leqslant{\sum_{j=1}^{N_S} x_{ij} \leqslant{1}},  \quad\forall i \label{const_1}\\
& x_{ij} \in \{0,1\}, \forall i; \forall j \label{const_5}\\
& x_{ij}d_{ij} + \underset{k=1}{\overset{i-1}\sum} q_{ijk} d_{kj} \leqslant {L_i}, \quad \forall i, \forall j \label{const_latency}\\
&  x_{ij} = 0  \rightarrow q_{ijk} = 0,   \forall i, \forall k\in \{1,...,i-1\};  \forall j  \label{const_3}\\
&  x_{ij} = 1  \rightarrow q_{ijk} = x_{kj},   \forall i, \forall k\in \{1,...,i-1\};  \forall j  \label{const_4}\\
& q_{ijk} \in \{0,1\}, \forall i, \forall k\in \{1,...,i-1\};  \forall j \label{const_6} 
\end{eqnarray}
where (\ref{const_1}) and (\ref{const_5}) say that a task can be assigned to at most one server. 
The three constraints (\ref{const_latency}), (\ref{const_3}), and 
(\ref{const_4}) collectively say that when assigned to server $j$, task $i$ should be completed no later than 
its deadline (i.e., the maximum allowed latency $L_i$).
Task $i$'s total delay on server $j$ is given 
by $x_{ij}d_{ij} + \underset{k=1}{\overset{i-1}\sum} q_{ijk} d_{kj}$, as shown in (\ref{const_latency}).
The two constraints (\ref{const_3}) and (\ref{const_4}) indicate that 
$q_{ijk}$ is equivalent to $x_{ij} x_{kj}$. Note that (\ref{const_3}) and (\ref{const_4})
are called indicator constraints in CPLEX solver \cite{cplex}. 
The $\underset{k=1}{\overset{i-1}\sum} q_{ijk} d_{kj}$ represents the queuing delay of task $i$
if it is assigned to server $j$. Recall that the tasks are served in a first-come first-serve order.
If task $k$ arriving before task $i$ (with $k \in \{1,...,i-1\}$) is also assigned to server $j$,
then task $i$ has to wait till task $k$ is finished. The queuing delay of 
task $i$ on server $j$ only makes sense when task $i$ is assigned to server $j$. 
Therefore, (\ref{const_3}) says that when task $i$ is not assigned to server $j$, 
its queuing delay constraint (\ref{const_latency}) on server $j$ should be removed.

\subsubsection{Dynamic task arrival}

The above optimization formulation for batch arrival of tasks illustrates the nature of the optimization problem 
to be solved by an Edge-Cloud system's manager, but it is difficult to implement in practice. This is 
because usually tasks arrive in a dynamic fashion, and since 
they have deadlines, they need to be sent to available servers immediately in order to meet their 
latency requirements.   

To address the case of dynamic task arrival, we introduce an online greedy algorithm 
(shown as Algorithm \ref{alg_greedy}) to be used by a system manager  
to maximize its total received revenue.
The idea of the algorithm is: whenever a server is available, it should be given
the task with the highest value among all tasks that are present in the system
and can be completed before their deadlines by the server.

If all tasks arrive at time $0$, then Algorithm \ref{alg_greedy} 
is essentially a heuristic to solve the optimization problem (\ref{val_max}) formulated for the batch task arrival case. 
In addition, if a task's value is inversely proportional to its deadline, then 
Algorithm \ref{alg_greedy} is a type of earliest-deadline-first scheduling algorithm, but without
preemptive scheduling. In an Edge-Cloud system,
a server cannot suspend the execution of a task in order to execute some other task with higher 
priority, due to the non-negligible communication cost/delay
in edge computing environment.
Both batch arrival and dynamic arrival of tasks will be investigated in Section \ref{sec_exp}.

\begin{algorithm}
	\caption{ \textbf{Online Greedy Task Distribution Algorithm}}
	\label{alg.mchain}
	\begin{algorithmic}[1]
		\REQUIRE $\langle \mathbb{N}_J(T), \mathbb{N}_S, T\rangle$, 
		where $T$ is the time period during which the algorithm executes,
		and $\mathbb{N}_J(T)$ is a set of tasks and their arrival times during $T$, and $\mathbb{N}_S$ is a server set.
		\STATE $t \leftarrow 0$, $Q =\emptyset $ ($Q$ is a priority queue where the task with the highest value is at the front of $Q$).
		
		\WHILE{$t \le T$} 
		\STATE If a task arrives at time $t$, insert it into $Q$:
		% where the task with the highest value is at the front of $Q$; 
		\begin{ALC@g}
			\STATE If multiple tasks have the same value, order them according to their arrival time order.  
		\end{ALC@g}
			
		\STATE If a set of servers are available at time $t$ (denoted by $\mathbb{S}_t \subseteq \mathbb{N}_S$), 
		use a loop to select all servers one at a time and in random order from $\mathbb{S}_t$, and for each selected server $svr_j$:
		% $svr_j$ in $\mathbb{N}_S$ is available at time $t$:
		\begin{ALC@g}
		\STATE Start from the front of $Q$, search for the task with the highest value
		among all tasks that can be finished 
		before their deadlines if processed by $svr_j$. Let $task^*$ denote such a task. 
		\begin{ALC@g}
		\STATE If $task^*$ is found, stop search and start a new thread for $svr_j$ to work on  $task^*$.
		%\STATE Otherwise, continue. 
		\end{ALC@g}
	\end{ALC@g}
		\ENDWHILE
	\end{algorithmic}\label{alg_greedy}
\end{algorithm}

\section{Mechanisms for revenue sharing}\label{sec_sharing}

In this section, we investigate the following three revenue sharing mechanisms: 
(1) Shapley value \cite{shapley}; 
(2) A proportional sharing mechanism proposed by Ortmann \cite{ortmann2000proportional}, 
referred to as \textit{Ortmann proportional sharing}; (3) and a sharing mechanism based
directly on each server's actual contribution, referred to as \textit{Direct-contribution-based} sharing mechanism.

\subsection{Shapley-value revenue sharing mechanism}

Shapley value \cite{shapley} is a well-known revenue sharing mechanism.
For an Edge-Cloud system, Shapley value defines
a function that distributes among a set of servers
the total revenue received by the system in organizing
the servers to work on a set of tasks.
It specifies that the revenue a server receives
equals the server's expected marginal contribution. 
 
Formally, consider a set of tasks $\mathbb{N}_J$, and a set of servers $\mathbb{N}_S$ (with $N_S=|\mathbb{N}_S|$). 
Note that a server can be owned by a cloud provider or an edge provider.
Define \textit{the value of set $\mathbb{S}$}, denoted by $v(\mathbb{S})$, as the total received value by only using servers 
in set $\mathbb{S}$ (with $\mathbb{S} \subseteq \mathbb{N}_S$) to work on the tasks in  $\mathbb{N}_J$. 
%Let $v(\mathbb{N}_J, \mathbb{N}_S)$ denote the total value received by the system through working on the set of tasks.
Note that $v$ is a function of task distribution mechanism.  
Let $\phi_{i}$ denote the revenue share given to server $i$.
The Shapley value mechanism 
assigns the following revenue share to server $i$:
\begin{equation} \label{equ_shapley}
\phi_{i}(\mathbb{N}_S) = \frac{1}{N_S!}\sum_{\mathbb{S}\subseteq{\mathbb{N}_S}\setminus\{i\}}|\mathbb{S}|!(N_S-|\mathbb{S}|-1)! \big(v (\mathbb{S}\cup\{i\})- v(\mathbb{S}) \big) 
\end{equation}
This revenue distribution mechanism satisfies the following desired property
\cite{shapley,myerson1977graphs,misra2010incentivizing,susanto2014ipccc}:
\textit{fairness or balanced contribution}, which says that 
for any pair of servers $i,j\in \mathbb{N}_S$, $j$'s contribution to $i$
equals $i$'s contribution to $j$, i.e., 
$\phi_i(\mathbb{N}_S) - \phi_i(\mathbb{N}_S\setminus\{j\} ) = \phi_j(\mathbb{N}_S) - \phi_j(\mathbb{N}_S\setminus\{i\} )$.
Shapley value sharing mechanism also has some other important properties such as 
efficiency (the sum of revenue shares distributed to all servers 
equals the total received revenue) and symmetry.

\subsubsection{Computing Shapley values}

The amount of time to compute Shapley values for all servers in a game is exponential,
if the computation is done according to the definition in Equation (\ref{equ_shapley}). 
However, we are able to derive a polynomial time algorithm, shown as Algorithm \ref{alg_shapley},
based on the following assumptions. 
The servers in a system can be divided into groups, which belong to  
different providers. For ease of exposition, assume that a provider has one and exactly one group.  
Further we assume that the servers in a group or provider are identical
(in terms of CPU, path bandwidth, and propagation delay)
as they are offered by the same provider. 
Then the Shapley values for all servers in a provider should be the same. 
Therefore for provider $k$, 
we just need to calculate a Shapley value $\phi_{i}$, where $i$
is an arbitrary server in the set of servers of provider $k$ (denoted by 
$\mathbb{N}_S^k$), i.e., $i\in \mathbb{N}_S^k$.
Then, provider $k$'s revenue is 
$v_k(N_k)=N_k \phi_i$, with $N_k=|\mathbb{N}_S^k|$.

\begin{algorithm}[h]
	\caption{ \textbf {Compute the Shapley value of each server of $m$ providers in an Edge-Cloud system.}}
	\label{alg.gm}
	\begin{algorithmic}[1]
		\REQUIRE Task set $\mathbb{N}_J$; Server set $\mathbb{N}_S =\bigcup_k \mathbb{N}_S^k$, where $\mathbb{N}_S^k$ is 
		the set of servers of provider $k$, $N_k= |\mathbb{N}_S^k|$, $N=|\mathbb{N}_S|$, $k=1,2,...,m$.
		\FOR{$j=1, 2, ..., m$}
		%\begin{ALC@g}
		\STATE Initialize $V^s_j=0$. 
		\FOR{$n_{j}=0$; $n_j \leq N_j-1$}
		\STATE Do a $m-1$ level nested loop to find all combinations 
		$(n_1, n_2, ..., n_{j-1}, n_{j+1}, ..., n_m)$,
		%(of the rest $m-1$ providers except $j$), 
		where each number $n_k$ (with $k\in \{1,2,..., j-1, j+1, ..., m\}$) varies from $0$ to $N_k$.
		\STATE For each $(n_1, n_2, ..., n_{j-1}, n_j, n_{j+1}, ..., n_m)$, do:
		\begin{ALC@g}
		\STATE Invoke a task distribution algorithm (e.g., \textbf{Algorithm \ref{alg_greedy}}) 
		for task set $\mathbb{N}_J$ to calculate two values $V_1$ and $V_2$ :
		\STATE $V_1 = V[n_1][n_2]...[n_{j-1}][n_j][n_{j+1}]...[n_m]$, i.e., the value of the set 
			that contains $n_k$ servers from provider $k$, with $k\in \{1,2,..., m\}$.
		\STATE $V_2 = V[n_1][n_2]...[n_{j-1}][n_j+1][n_{j+1}]...[n_m]$. (Similar to $V_1$, except that $n_j+1$ is used for provider $j$).
		\STATE Calculate $C_{coeff} =  C_{N_1}^{n_1} \cdot C_{N_2}^{n_2}\cdots C_{N_{j-1}}^{n_{j-1}}\cdot C_{{N_j}-1}^{n_j}\cdot C_{N_{j+1}}^{n_{j+1}} \cdots C_{N_m}^{n_m}$.
		(Note that $N_j-1$ is used for provider $j$).
		\STATE Let $S = \sum_{k=1}^m n_k$.
		\STATE Calculate $V^{inc}_{n_1, n_2, ... n_{j-1},n_j,n_{j+1}...n_m} = \frac{S!}{N!}(N-S-1)! \cdot C_{coeff} \cdot (V_2-V_1)$.
		\STATE Increase $V^s_j$ by $V^{inc}_{n_1, n_2, ... n_{j-1},n_j,n_{j+1}...n_m}$.
		\end{ALC@g}
		\ENDFOR
		\STATE Record $V^s_j$ (Shapley value of a server in provider $j$). 
		%\end{ALC@g}
	\ENDFOR
	\STATE Return Shapley value $\phi_{i\in \mathbb{N}_S^j}(\mathbb{N}_S)=V^s_j$,
	%where $\mathbb{N}_S^j$ denotes the set of servers of provider $j$,
	with $j={1, 2, ..., m}$. (All servers in a provider have
	the same Shapley value.)
	\end{algorithmic}\label{alg_shapley}
\end{algorithm}

\noindent\textbf{An example of computing Shapley values.}
We use a simple example of two providers to illustrate Algorithm \ref{alg_shapley}.
Consider a cloud provider that offers two cloud servers $c_1$, $c_2$ and an edge provider 
that offers three edge servers $e_1$, $e2$, and $e3$ in an Edge-Cloud system. 
According to equation (\ref{equ_shapley}), in order to derive the Shapley value of $c_1$, 
we need to find the values of sets $\mathbb{S}$ and $\mathbb{S}\bigcup\{c_1\}$. There are
$32$ such sets (including the empty set). Due to space limitations, they are not listed here.

However, all three edge servers can be treated as equivalent, and both cloud servers are also equivalent.
For example, the value of set $\{c_1, e1, e2\}$ equals the values of sets $\{c_1, e1, e3\}$
$\{c_1, e2, e3\}$, $\{c_2, e1, e2\}$, $\{c_2, e1, e3\}$ and $\{c_2, e2, e3\}$.
%, and then we can call those six sets are equivalent sets. 
Therefore, to calculate the Shapley value of cloud server $c_1$, we only need to calculate the values 
of the following \textit{eleven} sets $\{c_1\}$, $\{e_1\}$, $\{e_1, c_1\}$, $\{e_1, e_2\}$, $\{c_1, c_2\}$, $\{e_1, e_2, c_1\}$, $\{e_1, e_2, e_3\}$, 
$\{e_1, c_2, c_1\}$, $\{e_1, e_2, e_3, c_1\}$, $\{e_1, e_2, c_2, c_1\}$, $\{e_1, e_2, e_3, c_2, c_1\}$,
based on Algorithm \ref{alg_shapley}.

\subsubsection{Time complexity of Algorithm \ref{alg_shapley}}

Consider a system of two providers, in which 
an edge provider has a set of servers $\{e_1, e_2, \cdots, e_{N_1}\}$, 
and a cloud provider has a set of servers $\{c_1, c_2, \cdots, c_{N_2}\}$. All cloud servers are equivalent, 
and all edge servers are equivalent. Suppose we would like to apply Algorithm \ref{alg_shapley} to 
calculate the Shapley value of a particular edge server $e_i$. 
For any two subsets ${S_1}$ and ${S_2}$ that do not contain $e_i$, 
we have $v({S_1}) = v({S_2})$ and $v({S_1}\bigcup {e_i} )= v({S_2}\bigcup {e_i})$, 
if the number of edge servers in ${S_1}$ equals the number of edge servers in ${S_2}$, 
and the number of cloud servers in ${S_1}$ equals the number of cloud servers in ${S_2}$. 
Then, we only need to calculate the values of $(N_1+1)(N_2+1)-1$ sets. This is because, for a set listed in 
the Shapley value formula (\ref{equ_shapley}),  the set might contain a number of edge servers and the number 
can be  $0, 1, \cdots, N_1$; similarly, 
the set might contain a number of cloud servers and the number can be $0, 1, \cdots, N_2$. 
Thus the total number of the unique sets is $(N_1+1)(N_2+1)-1$, where the “-1” is for removing the value calculation 
for the empty set $\emptyset$ which is always zero. 
Therefore, the time complexity of Algorithm \ref{alg_shapley} is $O(N_1 N_2)$. 
In general, if there are $m$ providers which have $N_1, N_2, \dots, N_m$ servers, the time complexity of Algorithm \ref{alg_shapley} is $O(N_1  N_2\cdots N_m)$.

\subsection{Direct-contribution-based and Ortmann proportional sharing mechanisms}

The idea of Direct-contribution-based sharing is simple. A server is rewarded with a share of revenue 
that is proportional to the actual contribution that it has made
when working together with other servers to complete a set of tasks. 
In the case where a system manager distributes all of its received revenue among 
participating servers (i.e., with no revenue share left for itself), 
the amount of revenue that a server receives is exactly the same amount of payment given by the clients whose tasks are
completed by the server.  

Ortmann proportional sharing \cite{ortmann2000proportional} is a sharing mechanism that is similar to Shapley value, 
in the sense that it also relies on some calculation of the marginal contribution of a server, instead of
relying directly on the server's actual contribution.  
For example, for a system with only two servers $i$ and $j$, according to Ortmann proportional sharing, 
the revenue received by server $i$ should be 
$\phi_i^P=\frac{v(\{i\})}{v(\{i\}) + v(\{j\})} v(\{i,j\})$, 
where $v(\{ i \} )$ is the revenue or value generated by the system when it only contains server $i$, 
and $v(\{i,j\})$ is the revenue generated by the system when it contains servers $i$ and $j$.
For a formal definition of Ortmann proportional sharing, see \cite{ortmann2000proportional}. 

Note that Ortmann proportional sharing also has a 
balanced contribution property, but it differs from Shapley value's balanced contribution
in the following sense. Under Ortmann sharing, for any pair of servers
$i,j\in \mathbb{N}_S$, $j$'s contribution to $i$
equals $i$'s contribution to $j$ in terms of quotient, i.e., 
$\phi_i^P(\mathbb{N}_S) / \phi_i^P(\mathbb{N}_S\setminus\{j\} ) 
= \phi_j^P(\mathbb{N}_S) / \phi_j^P(\mathbb{N}_S\setminus\{i\} )$.
That is, Ortmann's balanced contribution property 
is in the form of ratio equality instead of the difference equality of Shapley's. 
%An simple example to compare Shapley value sharing with proportional sharing. 
For example, consider a system of two players $1$ and $2$. Let $v(\{1\})=2$, $v(\{2\})=6$, and $v(\{1,2\})=40$. 
Based on Shapley value sharing,
they will get $\phi_1^S=18$ and $\phi_2^S=22$; based on Ortmann proportional sharing, 
they will get $\phi_1^P=10$ and $\phi_2^P=30$.
The Shapley value sharing satisfies $18-2=22-6$, but Ortmann proportional sharing 
satisfies $10/2=30/6$.

\subsection{Equilibrium state of an Edge-Cloud system}

Recall that we model the competition among multiple service providers as a non-cooperative game, which 
is under the two system-level mechanisms: task distribution mechanism and revenue sharing mechanism. 
In this section, a theorem of the equilibrium state of this game is first proposed and then proved. \\
%we first show that the game is finite and then we 
%illustrate the equilibrium state of the system.
\textbf{Theorem 1} \textit{The game of multiple resource providers in an Edge-Cloud system has a Nash equilibrium point, which is 
	an equilibrium system state.} \\
\begin{proof}
	Recall that in an Edge-Cloud system there is a set of computing tasks denoted 
	by $\mathbb{N}_J$, a set of servers denoted by $\mathbb{N}_S$ with $N_S = |\mathbb{N}_S|$, and 
	$m$ providers. Without loss of generality, we consider provider $i$, and its set of servers is denoted by $\mathbb{N}^i_S$, 
	and its number of servers is denoted by 
	$N_i= |\mathbb{N}_S^i|$, $i=1,2,...,m$. 
	Let $-i$ represent the set of all providers except provider $i$, $V_{opt}(N_i)$ denote the maximum value of the objective function in Eqn. (\ref{val_max}).
	Let $N_{-i}$ be fixed, $V_{opt}(N_i)$ is a non-decreasing function of $N_i$. 
	
	There is a threhold of $N_i$, and let $\bar{N_i}$ denote this threhold. When $N_i \geq \bar{N_i}$ and $N_i$ increases, $V_{opt}(N_i)$ remains the same value. 
	That is, once there is a sufficient number of servers 
	from provider $i$ that can satisfy the server allocation to tasks to get the optimal value (in solving the task distribution optimization problem), then further increasing $N_i$ 
	will not get any additional value increase of the objective function. 
	
	Let $V_{opt}^{*}$ denote the stable maximum objective value of $V_{opt}(N_i)$, that is,
	
	\begin{comment}
	Suppose that we fix the numbers of the servers of all other providers. That is, $N_{-i}$ is fixed, 
	where $-i$ represents the set of all providers except provider $i$. 
	Let $V_{opt}(N_i)$ denote the maximum value of the objective function in Eqn. (\ref{val_max}). It is easy to know that $V_{opt}(N_i)$ is a non-decreasing function of $N_i$. 
	When we increase $N_i$, $V_{opt}(N_i)$ will remain to be the same once $N_i$ is greater than or equal to a specific number $\bar{N_i}$. 
	That is, once there is a sufficient number of servers 
	from provider $i$ that can satisfy the server allocation to tasks to get the optimal value (in solving the task distribution optimization problem), then further increasing $N_i$ 
	will not get any additional value increase of the objective function. 
	Let $V_{opt}^{*}$ denote the stable maximum objective value of $V_{opt}(N_i)$, that is,
	\end{comment}
	\begin{equation}
	V_{opt}^{*} = V_{opt}(\bar{N_i}) .
	\end{equation}
	Let $V_i(N_i)$ denote the revenue received by provider $i$ as the result of a revenue sharing mechanism (e.g., Shapley value) that 
	splits the total value/revenue by solving the task distribution optimization problem. Then, 
	\begin{equation}\label{sup_opt}
	V_{opt}^{*} = \sup_{N_i \in  \mathbb{Z}^{+} \cup \{0\}} V_i(N_i) .
	\end{equation}
	In addition, Eqn. (\ref{eqn_u_general}) shows that the utility of provider $i$ is: 
	\begin{equation}
	U_i(N_i) = V_i(N_i) - \alpha_{i}N_i .
	\end{equation}
	If we take Eqn. (\ref{sup_opt}) into consideration, then we have 
	\begin{equation}
	U_i(N_i) = V_i(N_i) - \alpha_{i}N_i \leqslant V_{opt}^{*} - \alpha_{i}N_i .
	\end{equation}
	Note that $\alpha_{i} > 0$, thus when $N_i > \frac{V_{opt}^{*}}{\alpha_{i}}$, $U_i(N_i) < 0$. In addition, 
	this also indicates that once $N_i$ reaches a certain value, provider $i$'s utility will be a decreasing 
	function of $N_i$. Therefore, $N_i$ is upper bounded\footnote{In practice, $N_i$ is also upper bounded as 
		the number of servers of a provider is always limited.}. Meanwhile, the number of servers of any provider 
	must be a non-negative integer, which means $N_i \in  \mathbb{Z}^{+} \cup \{0\}$. The above discussion shows that the game is finite \cite{noncooperativegame}. 
	Since every finite game has an equilibrium point \cite{noncooperativegame}, we can obtain the theorem. 
\end{proof}
\textbf{Remarks:} We observe in our experiments that pure-strategy 
Nash equilibrium also exists in some cases, which are shown in Figures \ref{best_resp} and \ref{best_response_d_4} in Section V.C. When there is no 
pure-strategy Nash equilibrium state, we use the system performance at mixed-strategy Nash equilibrium to represent system stable state. 
When we observe multiple Nash equilibria in some cases in our experiments, we examine the average system performance
at those equilibrium states.

\section{Impact of Revenue Sharing Mechanisms on System Performance}\label{sec_exp}

%\section{Experiments and simulations}\label{sec_exp}

We discuss in this section the impact of revenue sharing mechanisms on the performance 
of an Edge-Cloud system.
Our investigation is mainly conducted through 
a combination of emulations and simulations\footnote{Due to resource constraints, 
	it is impossible for us to conduct Internet-scale experiments. 
	Therefore we mainly rely on simulations with system parameters 
	derived from our experiments and empirical trace.}.

\subsection{Edge-Cloud emulation system}

\begin{figure}[htb!]
	\centerline{
		\begin{minipage}{1.3in}
			\begin{center}
				\setlength{\epsfxsize}{1.3in}
				\epsffile{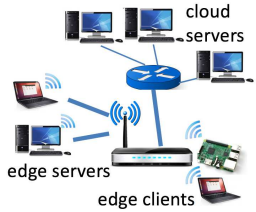}\\
				{}
			\end{center}
		\end{minipage}
	}
	\caption{Edge-Cloud emulation system.}
	\label{fig_emulation}
\end{figure}

We have built an experimental system to emulate an Edge-Cloud system, and 
based on which we have conducted experiments to derive system parameters to drive our simulations. 
The system consists of a pool of edge clients (on a number of Raspberry Pi's \cite{rasp} and Ubuntu laptops), 
a system manager (a distributed software component), and a pool of servers (on 
a number of Ubuntu workstations), as shown in Figure \ref{fig_emulation}.
A client at the edge submits its computation tasks to the manager at the edge who schedules and dispatches 
received tasks to servers.
There are two types of servers in the system: edge servers and cloud servers, 
which respectively belong to
edge service providers and cloud service providers.
%These two types of servers differ in their propagation delays 
%and the bandwidth of the routes between them and the edge clients. 
The edge servers have higher bandwidth
and shorter propagation delays (on the paths between them and clients) than the cloud servers.
Once a server receives a task, a Docker container \cite{docker} will be launched on the server
to process the task. Once the task is completed, the server will notify the manager and sends back 
the result of the task to the edge client. 
The communication between the clients, the manager, and the servers 
is through Web Application Messaging Protocol (WAMP) \cite{wamp}, a real-time messaging protocol.
%and Crossbar.io \cite{crossbar} router for routing messages.  
%Our system utilizes both Remote Procedure Calls and Publish/Subscribe messaging patterns in WAMP.
%In addition, the Crossbar.io \cite{crossbar} is used as a router for routing messages in our system.  

\subsection{Determining Simulation Parameters}\label{sec_para}

We utilize image processing (a typical edge computing application) in our simulations
to investigate the impact of revenue sharing mechanisms. In this subsection, 
we discuss how to derive system parameters used in our simulations.

We focus on an object detection application, i.e., 
a client's task is to detect whether 
a specific object appears in a collection of images. The client 
submits a collection of images including the image of the target object and 
a number of candidate images in a batch to the system, 
and then the system assigns the task to a server. 
The server launches a Docker container \cite{docker} 
to process the images using OpenCV \cite{bradski2008learning}. 
%Once the Docker container finishes
%the job, it will sends it's analysis result back to the client, and the resources allocated to 
%the container are returned to its hosting server's OS. 

A simulation run is characterized by a group of system parameters: 
\begin{equation}
\{T, \mathbb{N}_J(T), \mathbb{N}_S,\lambda, f_{delay}, k_{latency}, k_{bw}\}
\end{equation} 
where $T$ is the system time duration we simulate,  $\mathbb{N}_J(T)$ denotes the set of tasks
and their arrival times during $T$, $\mathbb{N}_S$ denotes the set of servers,
$\lambda$ denotes task arrival rate,
%(number of tasks per minute), 
$f_{delay}$ denotes the function to calculate the completion time of a task, 
$k_{latency}$ denotes a latency factor, and
$k_{bw}$ denotes a bandwidth factor. These parameters are discussed below.

In our simulations, we let the size of a task be a uniform random number in range $[1,20]$ MB.
We can think of this task size in 
the context of an object detection application as follows.
The average task size $10$ MB (in our simulations) roughly corresponds to a batch of $6$ images with a regular image size (about $1.6$ MB 
on a typical smartphone). 
The average task size also roughly corresponds to a collection of $63$ images from the 
Microsoft COCO image dataset \cite{lin2014microsoft} with an average image size of $159$ KB.
The value of a task is a number chosen uniformly at random from range $[1,5]$.

We assume that tasks arrive at the system in a Poisson process with rate
$\lambda$ (number of tasks per minute). Poisson process is a
typical stochastic process used in modeling job arrival process.
We choose $\lambda=40$ in our simulations, 
which is the average job arrival rate in Google cloud trace \cite{googleclouddata}.

The completion time $f_{delay}$ of a task on a server depends on the server's CPU and bandwidth (of 
the path between itself and clients).
Through our experiments on the emulation system, we find that for the object detection application, 
the computation time of processing a batch of images is linearly proportional 
to the size of the batch. For example,
we tested a server in our experiments, which was a Dell mobile workstation 
with Intel Core $i7$ $2.60$ GHz, $4$ cores CPU, and $16$ GB memory.
We randomly selected $n$ images from Microsoft COCO image database \cite{lin2014microsoft} 
and then ran the object detection application with OpenCV. 
Each experimental setting was repeated $10$ times. 
We choose $n=10, 50, 100, 200, 300$, and for each $n$ value and each experiment
replicate, we recorded the total size of the batch of images. 
Let $t_i$ (sec) denote the computation time  of processing a batch of images of $s_i$ MB. 
A plot of  $t_i$ vs. $s_i$ shows a strong linear relationship between them,
and a linear regression analysis shows that $t_i= 2.6 s_i$.
We will utilize this function to calculate the computation time of a task in our simulations
reported in the rest of this section.

A task $i$ has a latency requirement, denoted by $L_i$ 
(i.e., the maximum allowed delay).
It is determined as follows. Let $L_{i,avg}$ denote the 
amount of time to complete task $i$ on an average server 
(i.e., a server with an average CPU power and average bandwidth to clients in the system) 
without considering any queuing delay (i.e., waiting for other tasks to be completed on the same server). 
Note that $L_{i,avg}$ includes computation time and task transmission time. 
Assume that the average upload bandwidth of the paths from clients to servers is $B$ Mbps,
and assume that an average server has a CPU similar to the one used in our experiments described above.
Then, $L_{i,avg}=2.6 s_i + (8 s_i/B) + d_{prop}$ sec, 
where $s_i$ is task $i$'s size (MB), 
and the average propagation delay $d_{prop}$ is negligible compared with computation and transmission delays.   
In our experiments, we let the bandwidth from a client to an edge server be $24$ Mbps (i.e., a WiFi environment),
and then we let the bandwidth from a client to a cloud server be $24/k_{bw}$ Mbps, 
where bandwidth factor $k_{bw}=1,2,3$, and $4$
model the practice where a cloud server usually has lower bandwidth to clients than an edge server. 
In calculating $L_{i,avg}$, we do not consider the time to transfer results back to clients
as the sizes of the results are very small and negligible. 
Then, we let $L_i$ be a number chosen uniformly at random from range $[L_{i,avg}, k_{latency} L_{i,avg}]$,
where latency factor $k_{latency}\ge 1$. 
The rationale of choosing such a latency requirement is that, 
a client should not expect that its task to be completed earlier than 
what an average server can offer; and it is reasonable for a client  
to expect its task to be completed not $k_{latency}$ times longer than what an average server can offer. 
%We have looked different levels of $k_{latency}$, reported later. 
The actual deadline of task $i$
is given by $a_{i}+L_i$, where $a_i$ is the arrival time of task $i$.

Recall that the utility of a provider $p$ is given by 
$U_{p} = v_{p}(n_{p}) - \alpha_{p} n_{p}$.
We choose a value for $\alpha_{p}$ to make the cost of providing a certain number of servers be 
comparable to the revenue received due to making those servers available\footnote{As a future work, we will decide empirically 
the cost coefficient $\alpha_p$ by considering actual operating cost, e.g., power consumption, electricity usage.}. 
%We determine $\alpha_{p}$ as follows. 
Specifically, we simulated an Edge-Cloud system with an edge player and a cloud player, and with one-minute simulation time
and $\lambda=40$. We repeated the simulation for various combinations of the numbers of edge and cloud servers,
with a total number of servers varying from $2$ to $30$. Based on these simulation runs,
we choose $\alpha_{edge}=4$ and $\alpha_{cloud}=3$ for edge and cloud player respectively. 
We let $\alpha_{edge}> \alpha_{cloud}$, as typically a cloud provider can deploy servers with a lower cost
due to it's economy of scale compared with edge providers.

\subsection{Existence of Nash equilibrium and efficiency loss metric}

We have conducted extensive simulations of dynamic task arrivals via an event-driven simulator that we developed in Python
and we have applied CPLEX \cite{cplex} to solve the optimization problem for tasks arriving in a batch.
Our results have demonstrated the existence of Nash equilibrium.
%, defined in Equation (\ref{eqn_ne}).
In this subsection, we first 
illustrate the structure of the game between two players, 
through our results on the case of batch task arrivals, and  
%use a static batch-job setting and the one-shot optimization formulation (\ref{const_1}) 
we introduce a metric to measure the efficiency of an Edge-Cloud system.
Then we discuss the impact of revenue sharing
via simulations of dynamic task arrivals.

Recall that for a batch arrival of tasks, the solution of optimization problem (\ref{val_max}) 
gives a system manager the maximum total received value or revenue 
by optimally assigning those tasks to its servers,
and then the manager utilizes a revenue sharing mechanism to split the revenue among its servers.

\begin{figure}[htb!]
	\centerline{
		\begin{minipage}{1.6in}
			\begin{center}
				\setlength{\epsfxsize}{1.6in}
				\epsffile{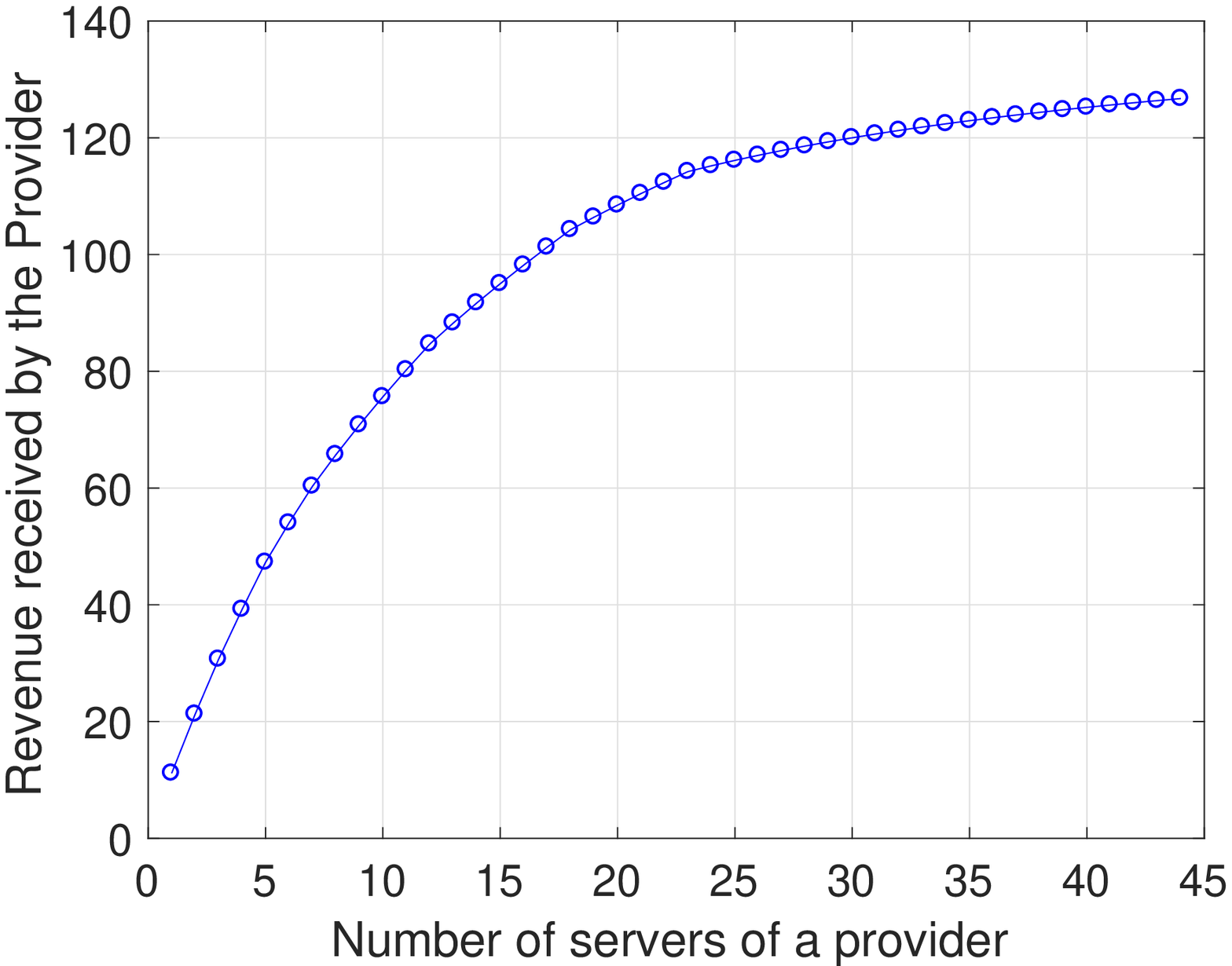}\\
				{}
			\end{center}
			\caption{Revenue of a provider.}
			% when it varies its number of servers.}
			\label{revenue_player2}
		\end{minipage}
		\begin{minipage}{1.6in}
			\begin{center}
				\setlength{\epsfxsize}{1.6in}
				\epsffile{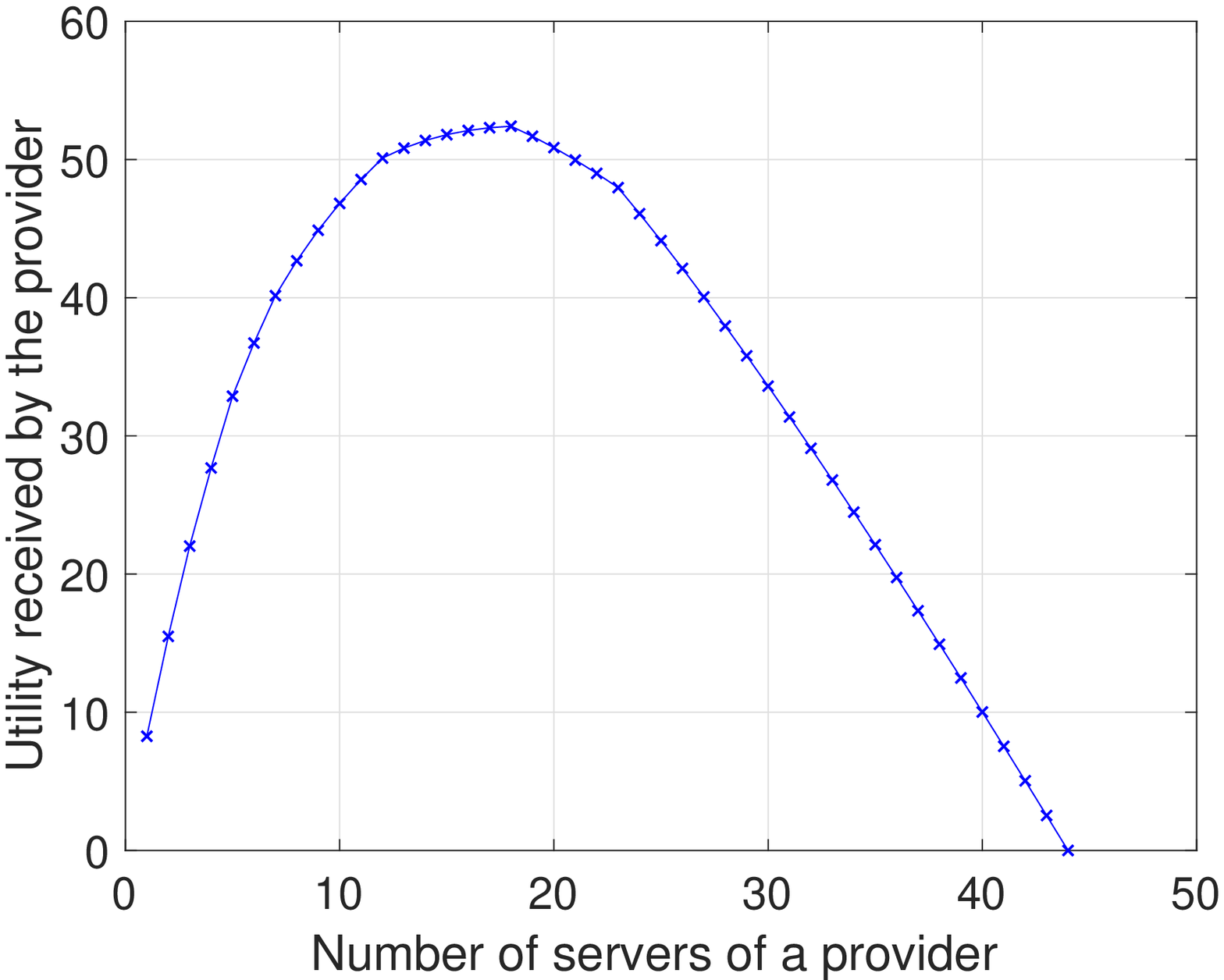}\\
				{}
			\end{center}
			\caption{Utility of a provider.}
			% when it varies its number of servers.}
			\label{utility_player2}
		\end{minipage}
	}
\end{figure}

\begin{figure}[htb!]
	\centerline{
		\begin{minipage}{1.6in}
			\begin{center}
				\setlength{\epsfxsize}{1.6in}
				\epsffile{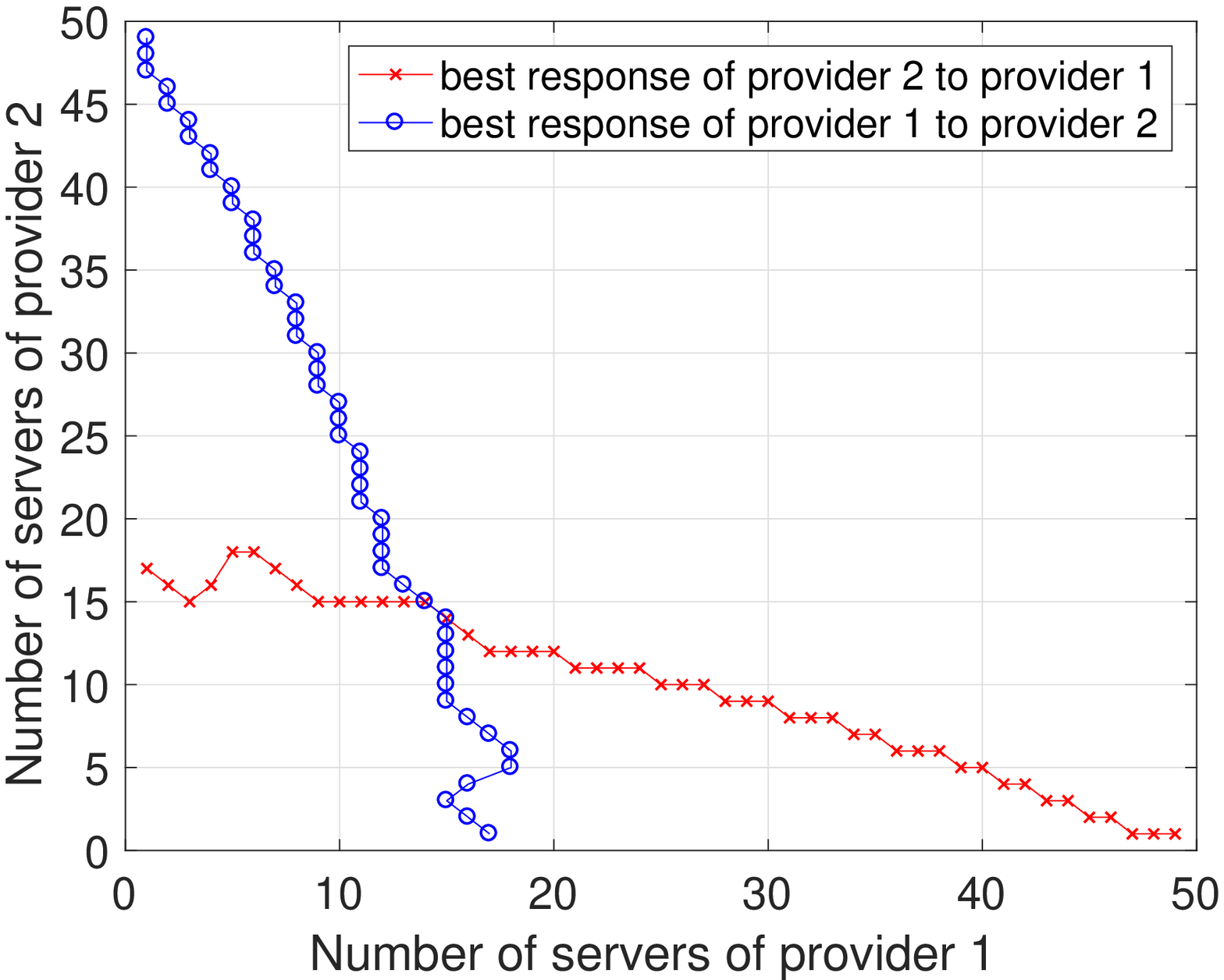}\\
				{}
			\end{center}
			\caption{Best response curves of two providers in a symmetric game.}
			% when they play a non-cooperative game together.}
			\label{best_resp}
		\end{minipage}
		\begin{minipage}{1.6in}
			\begin{center}
				\setlength{\epsfxsize}{1.6in}
				\epsffile{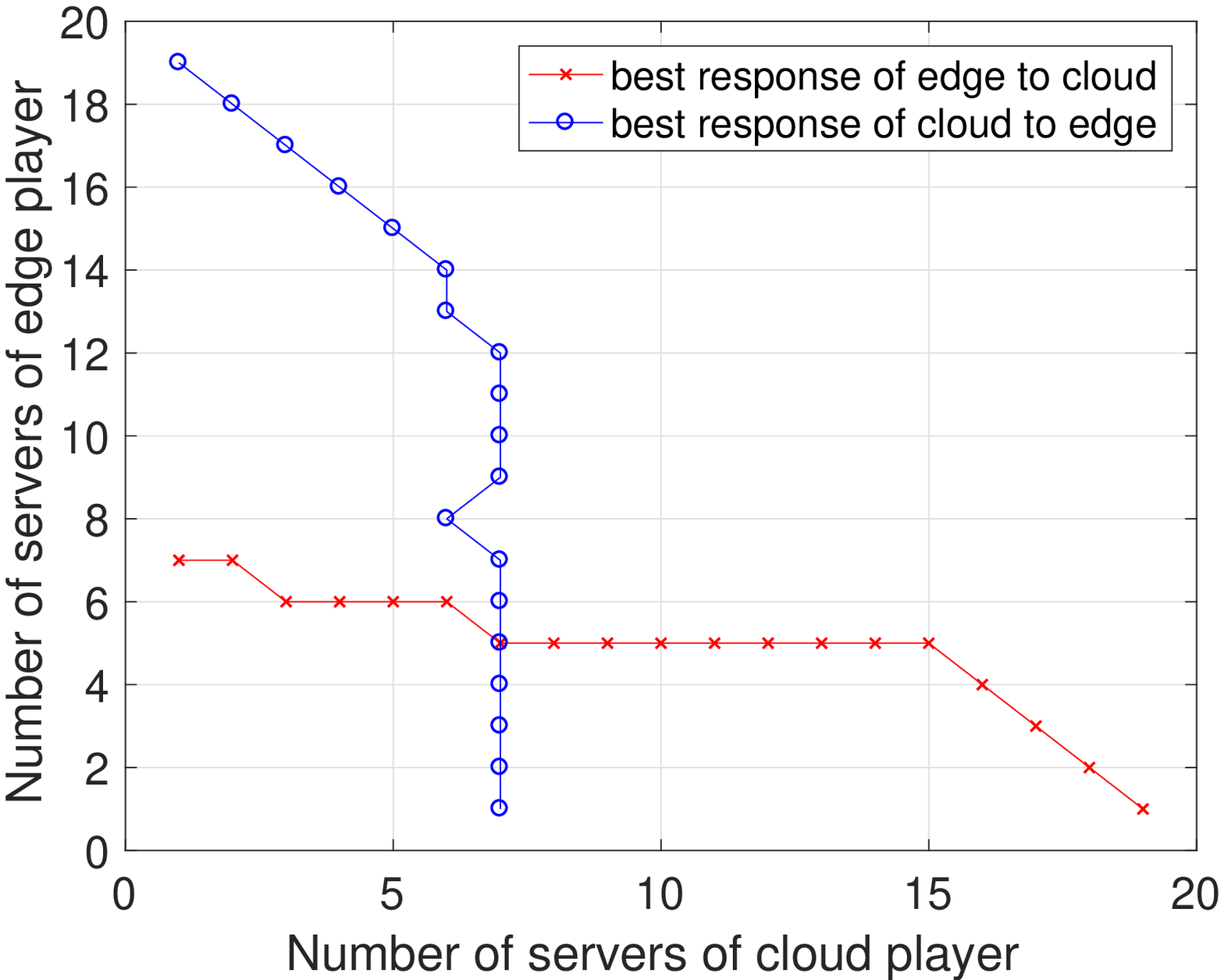}\\
				{}
			\end{center}
			\caption{Best responses of two providers in an asymmetric game.}
			% when the transmission bandwidth between clients and 
			%edge servers is four times faster than that between clients and cloud servers, i.e., $d=4$.}
			\label{best_response_d_4}
		\end{minipage}
	}
\end{figure}

\begin{figure*}[htb!]
	\centerline{
		\begin{minipage}{1.8in}
			\begin{center}
				\setlength{\epsfxsize}{1.8in}
				\epsffile{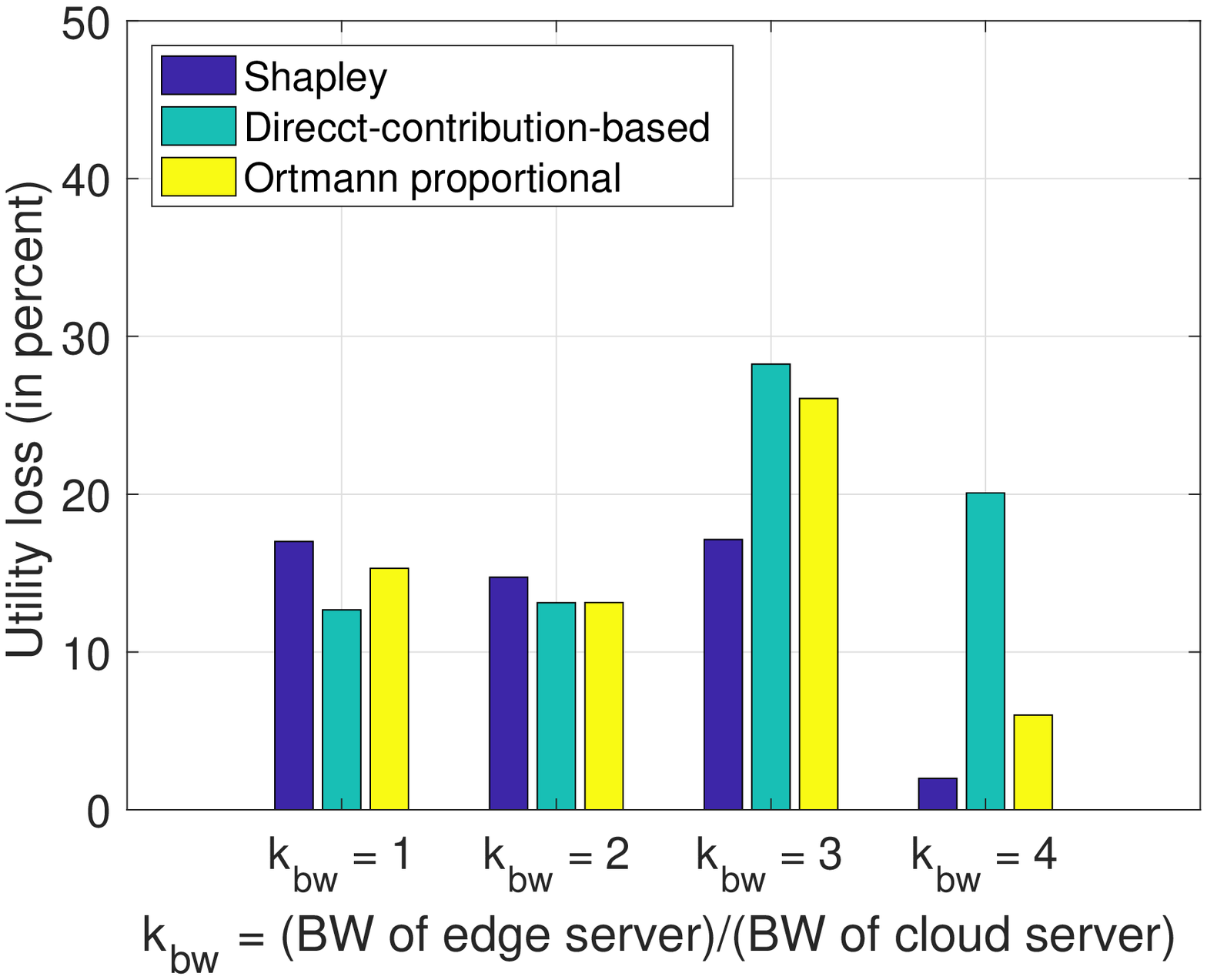}\\
				\small{(a) $k_{latency}=1.4$}
			\end{center}
		\end{minipage}
		\begin{minipage}{1.8in}
			\begin{center}
				\setlength{\epsfxsize}{1.8in}
				\epsffile{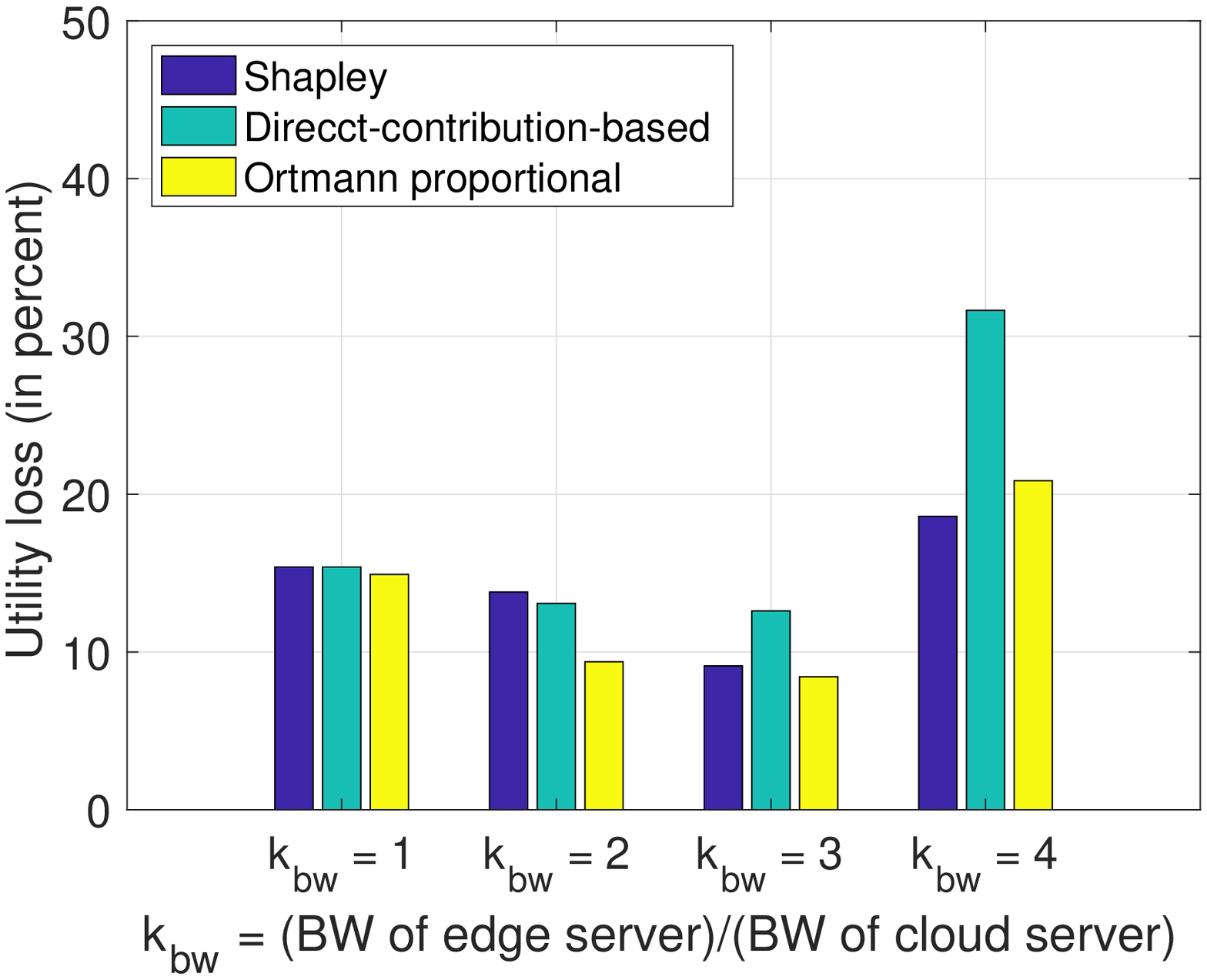}\\
				\small{(a) $k_{latency}=2$}
			\end{center}
		\end{minipage}
		\begin{minipage}{1.8in}
			\begin{center}
				\setlength{\epsfxsize}{1.8in}
				\epsffile{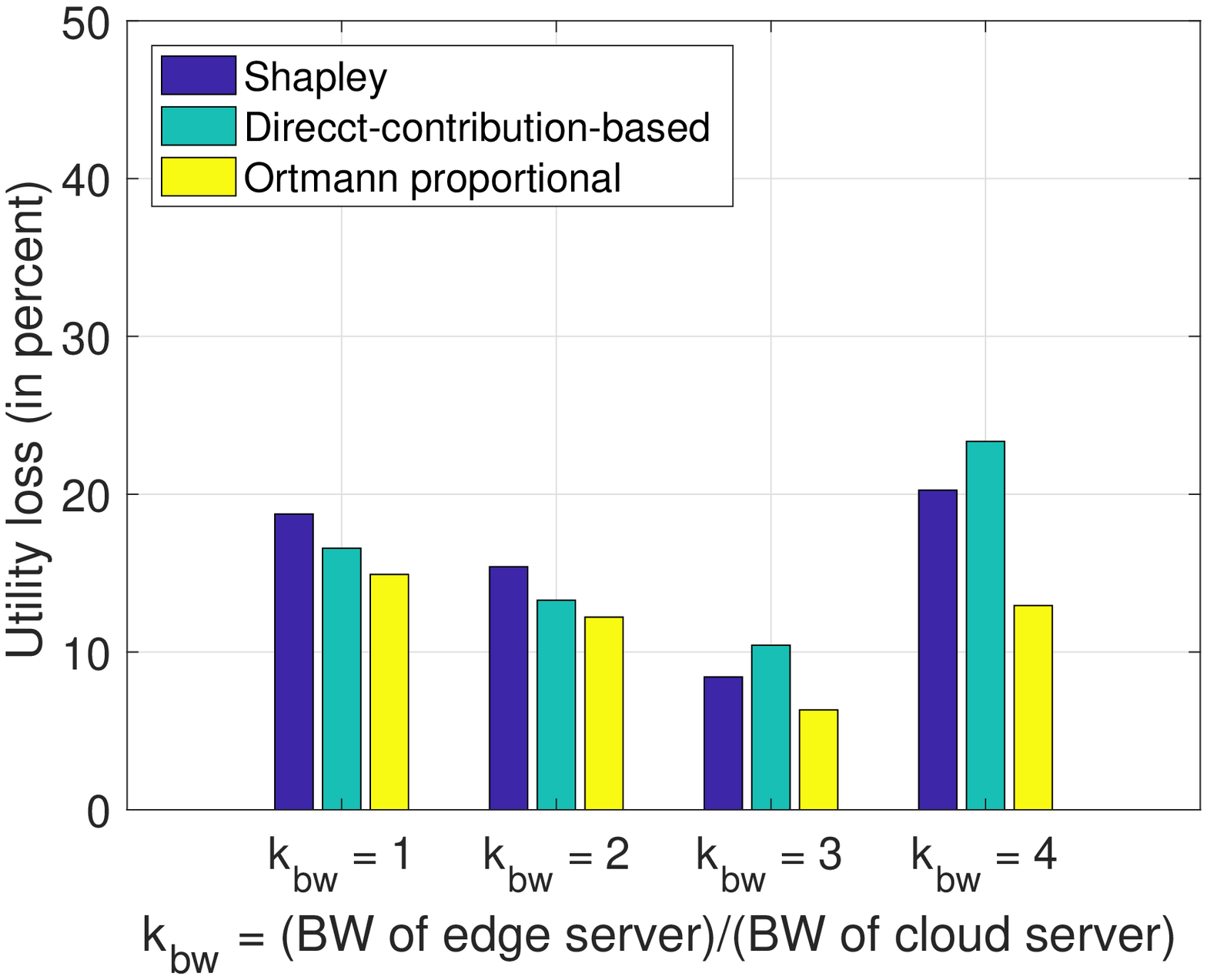}\\
				\small{(a) $k_{latency}=4$}
			\end{center}
		\end{minipage}
	}\caption{Utility loss (compared with the maximum utility) at Nash equilibria.}
	\label{utility_loss_compare_3_dynamic}
\end{figure*}
\begin{figure*}[htb!]
	\centerline{
		\begin{minipage}{1.8in}
			\begin{center}
				\setlength{\epsfxsize}{1.8in}
				\epsffile{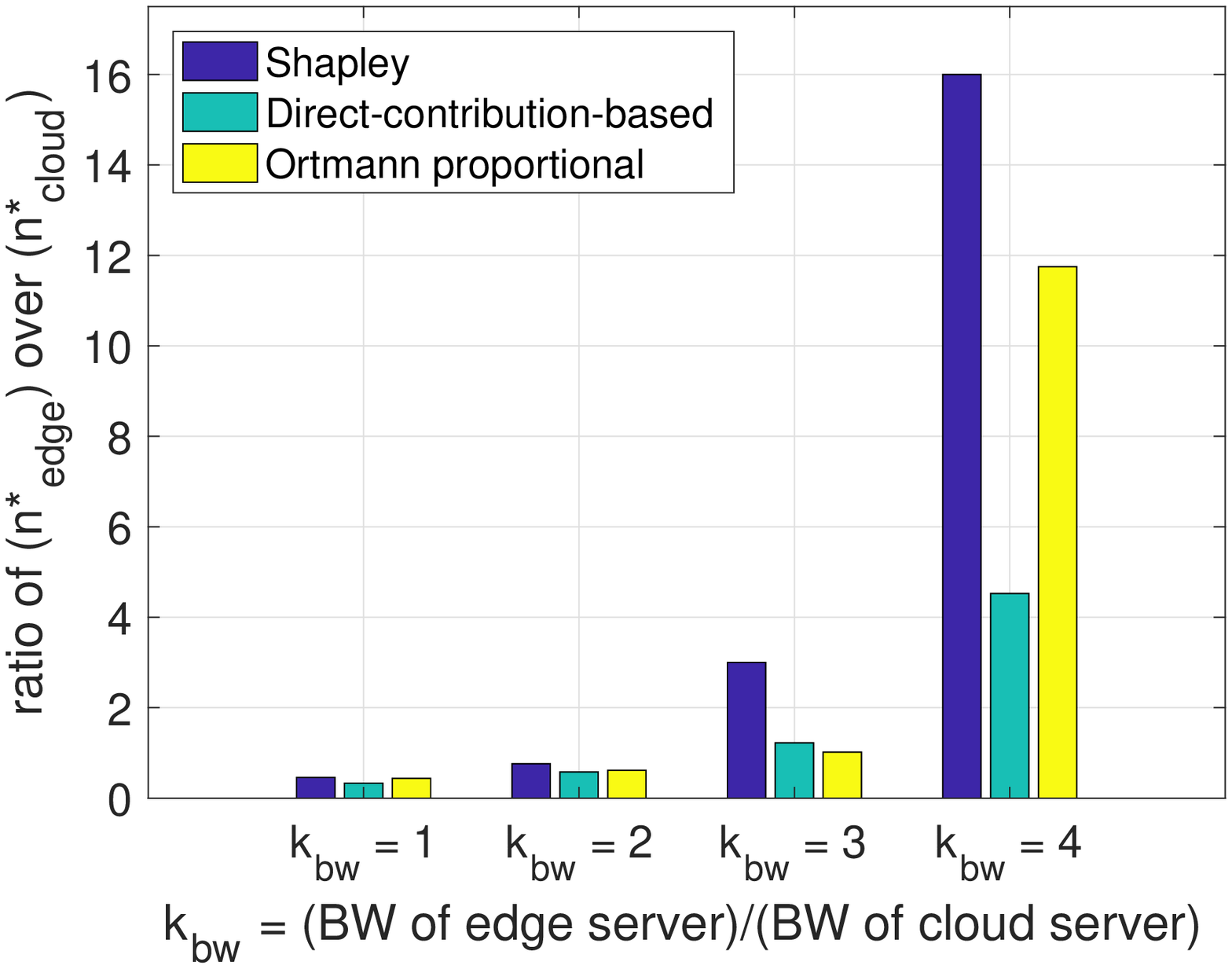}\\
				\small{(a) $k_{latency}=1.4$}
			\end{center}
		\end{minipage}
		\begin{minipage}{1.8in}
			\begin{center}
				\setlength{\epsfxsize}{1.8in}
				\epsffile{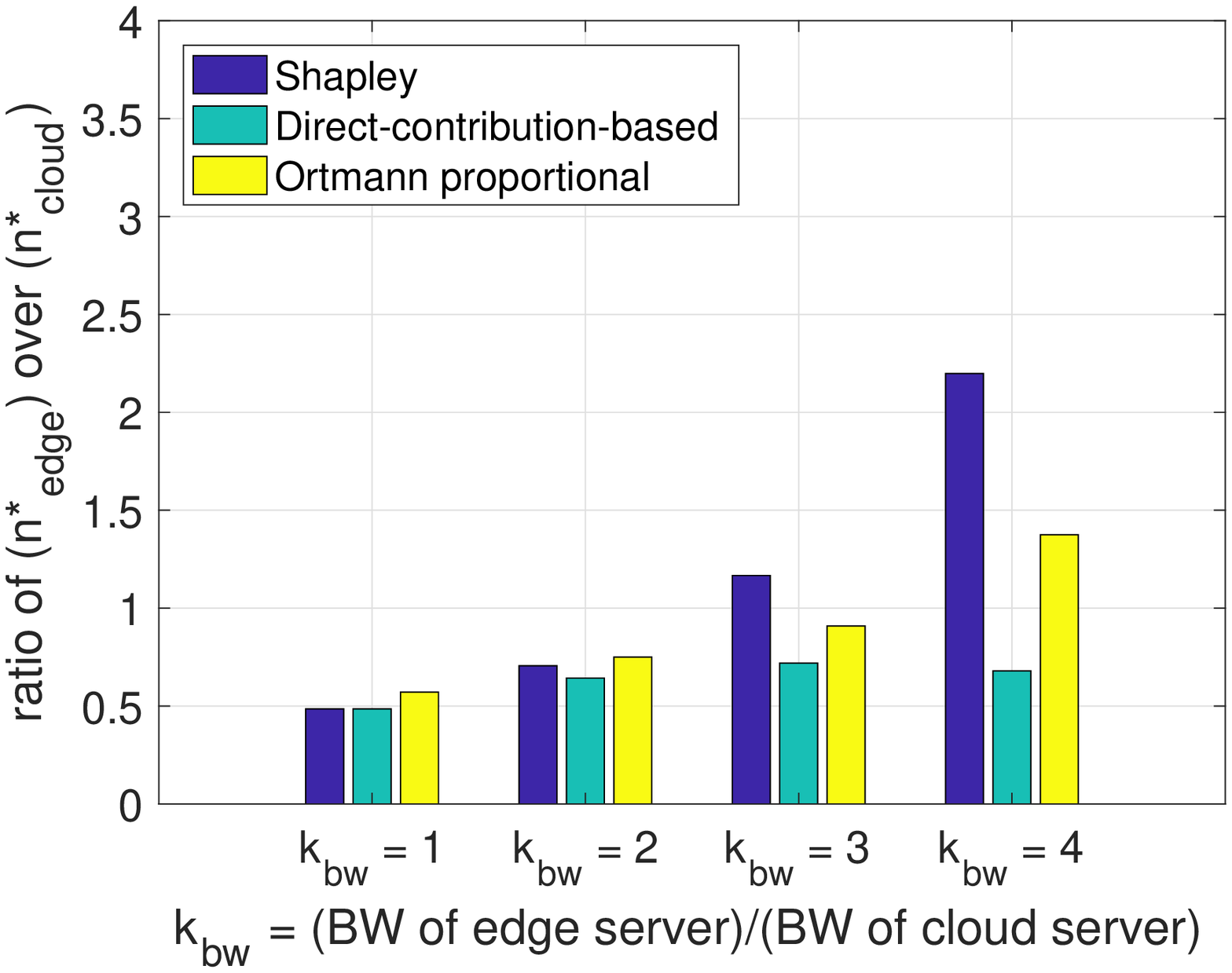}\\
				\small{(a) $k_{latency}=2$}
			\end{center}
		\end{minipage}
		\begin{minipage}{1.8in}
			\begin{center}
				\setlength{\epsfxsize}{1.8in}
				\epsffile{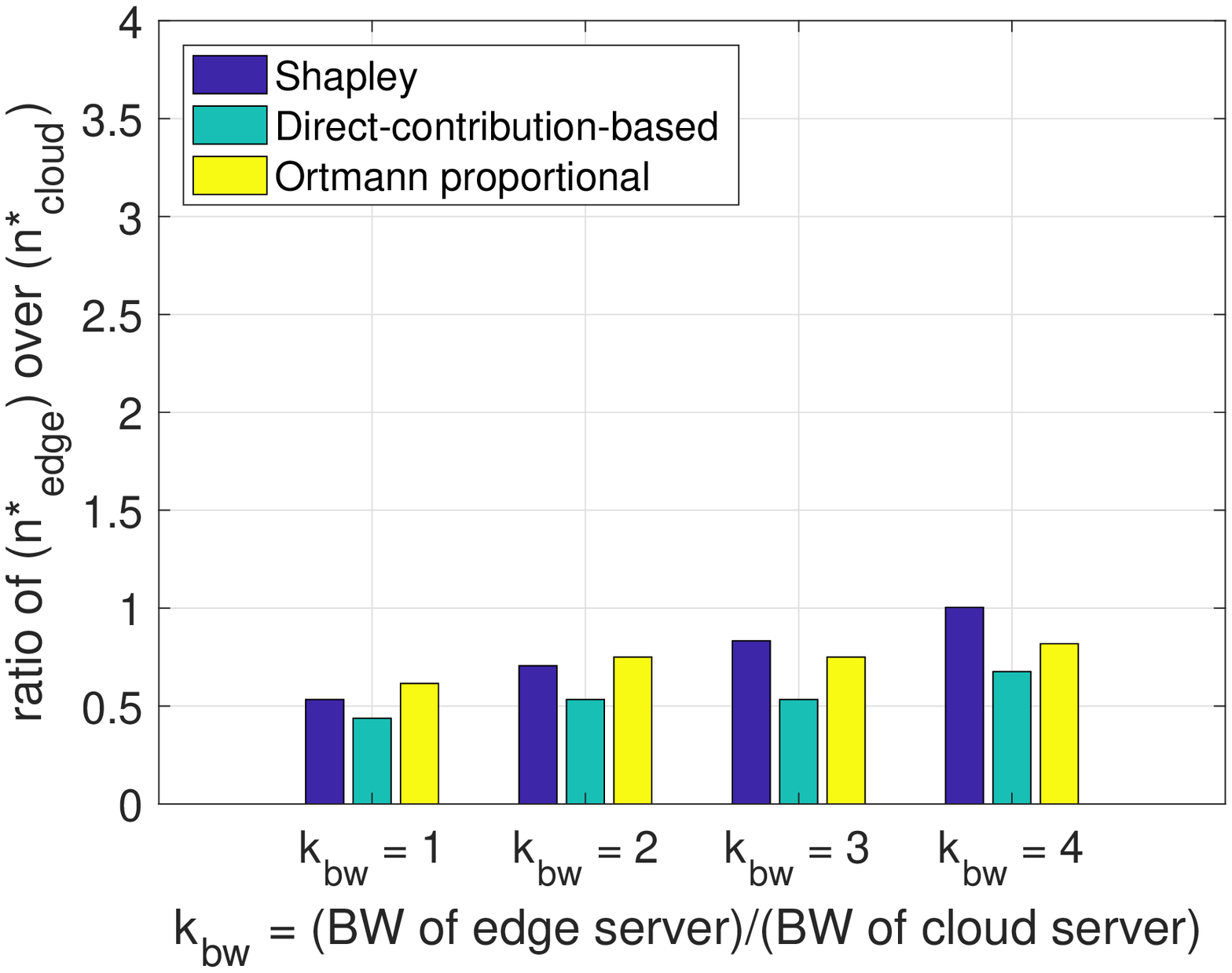}\\
				\small{(a) $k_{latency}=4$}
			\end{center}
		\end{minipage}
	}
	\caption{Ratio of the number of servers of the edge player over that of 
		the cloud player at Nash equilibria.}
	\label{ratio_d_together_dynamic}			
\end{figure*}

Consider a system that consists of two providers, with a set of tasks arriving in a batch, 
and Shapley value mechanism is adopted by the system.  
%(with a size distribution described in Subsection \ref{sec_para}). 
%We first consider a homogeneous environment where 
First suppose that the servers of both providers
are all identical in terms of CPU and bandwidth.
This game is referred to as a symmetric game.
Figure \ref{revenue_player2} shows the revenue received by a provider by varying its number of servers
in the system, when the other provider places $6$ servers in the system. 
The revenue curve is increasing and concave, which shows that the provider's marginal revenue  
is decreasing even though its received revenue is increasing due to its increasing number of servers. 
Figure \ref{utility_player2} shows the utility received by the provider.
% when it increases its number of servers placed in the system. 
We see that due to the decreasing marginal revenue
and the increasing cost, the utility first increases and then drops.
%Figure \ref{utility_total} shows the whole system's utility (i.e., the total utility of the two providers).

Based on the utilities of the two players, we derive their best response curves \cite{basar98game,zhang2005tcpconnection_icnp} 
and draw them in Figure \ref{best_resp}. 
Each point of a player's best response curve represents the player's best strategy in response to the other player's strategy.
For example, a point on the best response curve of player 2 to player 1 represents the 
number of servers (the y-axis value of the point) that gives player 2 the maximum utility given that player 1 chooses 
a certain number of servers (the x-axis value of the point).
Therefore, any intersection point of the two best response curves represents a Nash equilibrium. 
Figure \ref{best_resp}
shows that the game has two Nash equilibria  
$(n_1^*, n_2^*)=(14,15)$ or $(n_1^*, n_2^*)=(15,14)$, 
where $n_1^*$ is the optimal strategy of player $1$ with respect to player $2$'s strategy $n_2^*$, and vice versa.

Note that a typical metric to measure a system's performance at a Nash equilibrium is 
efficiency loss \cite{johari2004efficiency,zhang2005tcpconnection_icnp}, which is 
a comparison between the overall system utility at the equilibrium with the maximum
overall system utility. 
We use the relative \textit{utility loss} of an equilibrium
to capture the efficiency loss of the equilibrium, which is defined as 
\begin{equation}
U_{loss}=\frac{\big(\max_{\{n_p\}} \sum_{p} U_p(n_p)\big)- \big(\sum_{p} U_p^{NE}(n_p^{NE})\big) }{ \max_{\{n_p\}} \sum_{p} U_p(n_p) }
\end{equation}
where $U_p^{NE}$ is the utility of player $p$ at Nash equilibrium $NE$, 
and $\max_{\{n_p\} } \sum_{p} U_p(n_p) $ is the maximum overall system utility\footnote{It is 
	the solution of a system-wide utility maximization problem, 
	and the numbers of servers specified by the solution for the providers
	may not maximize their individual utilities.}.
For example, in the above game with $50$ tasks arriving in a batch,
we observe that the system's utility loss is around $19.5\%$ 
at the unique equilibrium\footnote{In some games,
	there are two Nash equilibria and they are very close to each other, similar to the case shown 
	in Figure \ref{best_resp}. The existence of multiple equilibria is due to the 
	discrete nature of strategies (i.e., number of servers). In those cases,
	utility loss is calculated as the average of those equilibria.}. 
As another example, we change the previous game
by setting $k_{bw}=4$ (then the game is asymmetric as the two players have different bandwidth).
Figure \ref{best_response_d_4} shows the existence of Nash equilibrium 
%(i.e., the intersection point of the best response curves) 
in this game.

%\subsection{Dynamic Arrival of Computing Tasks}

\subsection{Impact of Revenue Sharing on System Performance}

Our extensive simulations have demonstrated the existence of Nash equilibrium 
in the game between edge and cloud providers
in a wide range of system/network settings, when tasks arrive in a dynamic process.
We find that \textit{different revenue sharing mechanisms have quite different 
impacts on the performance of an Edge-Cloud system}, 
and in general \textit{Direct-contribution-based sharing mechanism results in the worst system-level efficiency
than Shapley and Ortmann mechanisms}. 
Due to space limitations, we only present in this section some results of the game between a cloud player 
and an edge player, shown in Figures \ref{utility_loss_compare_3_dynamic}
and \ref{ratio_d_together_dynamic}. 
For ease of exposition, we let the cloud player's servers differ from the edge player's servers only 
in the bandwidth of the paths between themselves and clients, and we let all servers have the same CPU capacity.
Each simulation presented here lasts $T=60$ minutes,
and tasks arrive at the system in a Poisson process with $\lambda=40$ tasks per minute. 
The results of our simulations based on the empirical task arrival process in Google trace \cite{googleclouddata}
are similar to the results presented in this subsection. 

\subsubsection{Utility loss}
We observe from Figure \ref{utility_loss_compare_3_dynamic} that 
when the transmission bandwidth difference between cloud servers and edge servers is small 
(i.e., $k_{bw}=1$ or $2$),
the losses of system-level utility are roughly the same 
for all three different revenue sharing mechanisms and across different 
latency requirement levels of tasks ($k_{latency}=1.4, 2, 4$)\footnote{If $k_{latency}=1.4$, then  $1.4 L_{i,avg}$
	is the amount of time to complete task $i$ (without queuing delay) on the server with the lowest bandwidth.}.

However, when the cloud player's transmission bandwidth is significantly lower than that of the edge player ($k_{bw}=4$),
Figure \ref{utility_loss_compare_3_dynamic} shows that Direct-contribution-based sharing gives
the worst utility loss among all three mechanisms, across different latency requirement levels ($k_{latency}$) of tasks.
In addition, when tasks have very stringent latency requirement 
(i.e., $k_{latency}=1.4$), Shapley mechanism gives the lowest utility loss
(only $1.98\%$).  
When latency requirement becomes less stringent ($k_{latency}=2$), 
Shapley mechanism and Ortmann mechanism have roughly the same utility loss. 
When tasks have very flexible latency requirements ($k_{latency}=4$), the utility loss 
of Shapley mechanism is higher than that of Ortmann mechanism, 
but still they both are lower than that of Direct-contribution-based mechanism. 
%Since $k_{latency}=2$ is more likely to represent a normal operating environment,  
%therefore Shapley value is the best sharing mechanism. 

\subsubsection{Numbers of servers at equilibria}
In addition, we have also examined the ratio of the number of the edge player's servers over the 
number of the cloud player's servers at 
Nash equilibrium\footnote{In the case where there are two Nash equilibria, 
	the ratio is calculated as the average of the 
	two equilibria, similar to the calculation of utility loss.}.
%Let $r*=n^*_{edge} / n^*_{cloud}$ denote the ratio. 
Figure \ref{ratio_d_together_dynamic} shows that when the bandwidth difference between 
the two players is not big ($k_{bw}=1$ or $2$),
the edge player has fewer number of servers at equilibria than the cloud player across all three different
revenue sharing mechanisms
and all three different levels of latency requirements. 
This is because the edge player's cost of placing servers in the system 
is higher than that of the cloud player.

When $k_{bw}$ gets higher ($k_{bw}=4$) and task latency requirement is stringent or normal ($k_{latency}=1.4$ or $2$), 
the cloud player will place fewer number of servers (than the edge player) at equilibria
under Shapley or Ortmann mechanism. This is
because the cloud player's servers are less likely able to meet tasks' deadline requirements, 
and the two revenue sharing mechanisms discourage the cloud player from putting more servers in the competition.
%(as those servers have much lower bandwidth than those of the cloud player).  
This discouragement leads to a better system-level performance (i.e., 
low system utility loss) than the Direct-contribution-based sharing, 
as shown in Figure \ref{utility_loss_compare_3_dynamic} (a) and (b).

\subsubsection{The case of large bandwidth difference ($k_{bw}=4$) 
	and stringent latency requirements of tasks ($k_{latency}=1.4$)}
The disadvantage of Direct-contribution-based sharing mechanism 
is quite obvious in this case.
% where tasks have very stringent latency requirements ($k_{latency}=1.4$)
%and cloud servers' bandwidth to clients are very low compared with that of edge servers ($k_{bw}=4$).
Figure \ref{utility_loss_compare_3_dynamic} (a) shows that 
its utility loss ($20.08\%$) is significantly higher than that of 
Shapley's ($1.98\%$) and Ortmann's ($6\%$). 

This is because that under the Direct-contribution-based sharing, the very low bandwidth provider (i.e., the cloud player) 
still aggressively utilizes many servers 
%(but they should not do so when they have very poor transmission bandwidth) 
in order to gain a high revenue (as they are rewarded directly based on their actual contributions), 
which leads to a very low overall system efficiency. 
But Shapley and Ortmann mechanisms discourage such an aggressive behavior of the provider
with very low bandwidth,
because our results show that in this case, Shapley and Ortmann mechanisms give
penalty instead of reward to the low bandwidth cloud servers.
Specifically, Shapley mechanism 
will start to assign decreasing or even negative revenue to cloud servers 
once the number of cloud servers increases over a threshold
(which implies that the addition of a cloud server with very low bandwidth will 
bring negative marginal contribution to the system);
and similarly, Ortmann mechanism will assign lower and lower revenues to cloud servers.
Our results also show that in this case,
Shapley and Ortmann mechanisms bring significant more revenue to the system 
than Direct-contribution-based sharing at Nash equilibria.
Our finding suggests that Shapley and Ortmann mechanisms, the two mechanisms 
that distribute revenue based on marginal 
contributions instead of directly on actual contributions, 
can help improve a system's overall utility
in the face of the self-interested behavior of providers.  

\noindent \textbf{Summary.} Shapley mechanism gives the best system-level performance 
in most cases in our simulations, despite that 
providers game with the system to pursue their self-interested optimization goals. 
Ortmann mechanism is similar to Shapley mechanism in terms of system utility loss at Nash equilibria.
Direct-contribution-based mechanism results in the worst system performance at equilibria. 

\section{Conclusions and Future work}\label{sec_conclusion}

We have proposed a game-theoretic framework to investigate the impact of revenue sharing mechanisms
on the performance of an Edge-Cloud system in which edge providers and cloud providers compete with each other 
and game with the system in order to maximize their own utilities. We have found that the revenue sharing based 
directly on actual contributions of servers can result in significantly worse system-level performance
than Shapley value and Ortmann proportional sharing mechanisms at the Nash equilibria of the game between providers. 
For future work, we will conduct a theoretic analysis, study dynamic game playing processes,
and conduct large scale experiments of Edge-Cloud systems.

\bibliographystyle{IEEEtran}
\bibliography{social_swarms}

\end{document}